\input psfig.tex
\documentstyle[prd,aps]{revtex}
\bibstyle{unsrt}
\tighten
\begin{document}
\draft
\preprint{Imperial/TP/94-95/49}
\twocolumn[\hsize\textwidth\columnwidth\hsize\csname
@twocolumnfalse\endcsname
\title{An Extension to Models for Cosmic String Formation}
\author{Andrew Yates\footnote{email: \tt{a.yates@ic.ac.uk}}
 and T.W.B. Kibble\footnote{email: \tt{t.kibble@ic.ac.uk}}}
\address{Blackett Laboratory, Imperial College, South Kensington,
London SW7 2BZ, U.K.}
\date{August 15, 1995}
\maketitle
\begin{abstract}
The canonical Monte-Carlo algorithm for simulating the production of
string-like topological defects at a phase transition is extended by
introducing a distribution of domain sizes. A strong correlation is
found between the fraction in the form of `infinite' string and the
variance of the volume of the regions of constant phase.
\end{abstract}
\vskip2pc]
%\pacs{PACS numbers : 98.80C}

\section{Introduction}
Cosmic strings may have formed at a phase transition in the early
universe \cite{Tom,Review}. Information about the initial statistics of
a string network, after the point at which thermal fluctuations
become unimportant and the strings are `frozen in', has largely emerged
from the numerical simulations first performed by Vachaspati and
Vilenkin \cite{VV}.
The simplest case, involving the spontaneous
breaking of a $U(1)$ symmetry, is mimicked by assigning a phase
between 0 and $2\pi$ to each point on a regular lattice.
The lattice spacing then
corresponds to a correlation length $\xi$ characteristic of the
scalar field acquiring the non-zero vacuum expectation value. To look for
field configurations with non-trivial topology, the `geodesic rule'
is invoked. This proposes that to minimise gradient energy the field
will follow geodesic paths on the vacuum manifold as a path in
configuration space is traversed. The phase will thus follow the
`shortest path' between values on adjacent lattice sites. A winding of
$\pm 2\pi$ around a plaquette in the lattice means that a line-like
distribution of zeroes of the field will pierce it --- a cosmic string.

If we impose periodic boundary conditions on our lattice, it becomes
obvious that all string must be in the form of closed loops. One would
expect that this procedure gives a string configuration in our box
that is statistically similar to that when
neighbouring, causally disconnected regions are present\cite{Ray}.
{}From the distribution of lengths of loops, it is easy to make
a distinction between `infinite' string (winding around the box many
times) and smaller loops, peaked at the minimum size of four lattice
spacings. The analytic form of the small-loop distribution is well
understood statistically. It is found that, for a cubic lattice,
around $70-80\%$ of string
exists as infinite string in this scenario, which has long been used
as the generator of initial configurations for the numerical evolution
of string networks \cite{AS,AT,BB}.

If a non-minimal discretisation of the vacuum manifold is used (that
is, in the case of a broken $U(1)$ symmetry, approximating $S^1$ with
more than the smallest number of points, $\theta =$ 0, $2\pi /3$,
$4\pi /3$)
and we employ a cubic
lattice of points, in principle it is possible for all six faces of a
fundamental cell to contain strings. Even in the minimal case, it is
possible for four faces to do so. This requires a random choice
to be made, pairing the incoming and outgoing strings. The only method
that avoids this ambiguity is to use a tetrahedral lattice with a
minimal discretisation, so that at most one string enters and leaves
each cell. String configurations arising from this model have been analysed
recently \cite{HS} and it is found that a slightly lower fraction
(around $65\%$) exists as infinite string.

As a string network evolves, by means of intercommutation and
expansion of the universe, it has been predicted and (to different
extents) observed in the simulations that the characteristic lengths
describing it approach a `scaling regime', in which they grow in
proportion to the horizon size. A typical evolving network will
display an initial flurry of loop production before settling into this
scaling regime with a few long strings and large loops per horizon
volume continuing to (self-) intersect and produce smaller loops.
As such, it has been supposed that the
initial details of the string network are largely washed out after a
few expansion times. This indeed seems to be the case, from both the
numerical work and more recent analytic models of network evolution
\cite{BigTom}.
The question arises, though --- is there a
causal mechanism for creating a string distribution with significantly
less infinite string? In the most extreme case, it is possible that if
there were no infinite string at all, all the loops would disappear
within a finite (and quite short) time.
Recent work by Ferreira and Turok \cite{Ped}
partly confirms this, showing that a different type of scaling occurs
in that case. One way of testing this idea is to
attempt to rectify one of the major simplifications inherent in the
Vachaspati-Vilenkin algorithm, and introduce a {\it
distribution} of domain volumes in the initial conditions, instead of
simply assuming that causally disconnected regions of one value of the
field are of equal volume ($\sim \xi^3$).

\section{Implementing the algorithm}
A cubic lattice was used with a near-continuous \footnote{
i.e. a very high-density discretisation of the circle, that allows
the use of integer arithmetic} representation of
the vacuum manifold, despite the reduction in ambiguity that can
be achieved with the tetrahedral lattice, as mentioned above, since
it simplified the process of creating a domain structure. Physical
space is partitioned into regions of constant $U(1)$ phase by
throwing down domains of random diameter within specified limits and
gradually covering the lattice, dealing with the overlap and fragmentation
of these regions as the box becomes filled with the broken phase.
Roughly spherical domains were experimented with at first. However,
after taking account of the significant domain overlap that resulted from
the random filling of the lattice, and to make the task of ensuring
that no domain was created entirely within another more
straightforward, cubical domains were used. Once this was completed,
strings were located and traced through the lattice, following the
edges of either three or four adjacent domains.

It should be made clear that this is no more than a means of setting up a
domain structure, and in no way claims to simulate the dynamics of an
actual phase transition. Indeed, it is not obvious what order of
transition the results of this algorithm apply to, though it
would seem more closely related to string
formation at the interfaces of expanding bubbles of the true vacuum,
rather than
the uniform emergence of a domain structure in a second-order transition.
As a first guess we might expect a Gaussian distribution of domain
volumes, peaked around some mean value. As it happens this is
difficult to realise, and the size distribution appears to be more
Poissonian
(Fig.\ 1). The results are interesting nevertheless, and in
particular the fact that the resulting form of the graph seems largely
insensitive to modifications to the domain-laying algorithm.

The range of sizes of domains laid down was
systematically increased in order to plot the fraction of the total string
density as `infinite' string, $f_{\infty}$, against
domain volume variance. The variance is normalised to the mean domain
volume in order to remove effects due to uniform scaling-up of
domain volumes.
In the zero-variance limit the Vachaspati-Vilenkin result of
$f_{\infty}\simeq 0.76$ is obtained. The precise value is weakly
dependent on the imposed loop/`infinite' string cutoff --- if we set the
maximum size of a loop to be $4N$, $10N$ and $N^2 /2$, where $N$ is the
length of the side of our box in units of the smallest possible domain
size, the zero-variance values of $f_{\infty}$ are 0.78, 0.77 and
0.75 respectively. Higher-variance values of $f_{\infty}$
change by a similar amount. A cutoff of $N^2 /4$ was used in the
plots presented here.

%%%%%%%%%%%%%%%%%%%%%%%%%%%%%%%%%%%%%%%%%%%%%%%%%%
% Histograms figures here
%%%%%%%%%%%%%%%%%%%%%%%%%%%%%%%%%%%%%%%%%%%%%%%%%%
% Two col. format
\begin{figure*}
\label{fig:hist}
\begin{minipage}{4.5in}
\setlength{\unitlength}{1in}
\begin{picture}(2,4.5)
\put(0.5,0.40){\psfig{file=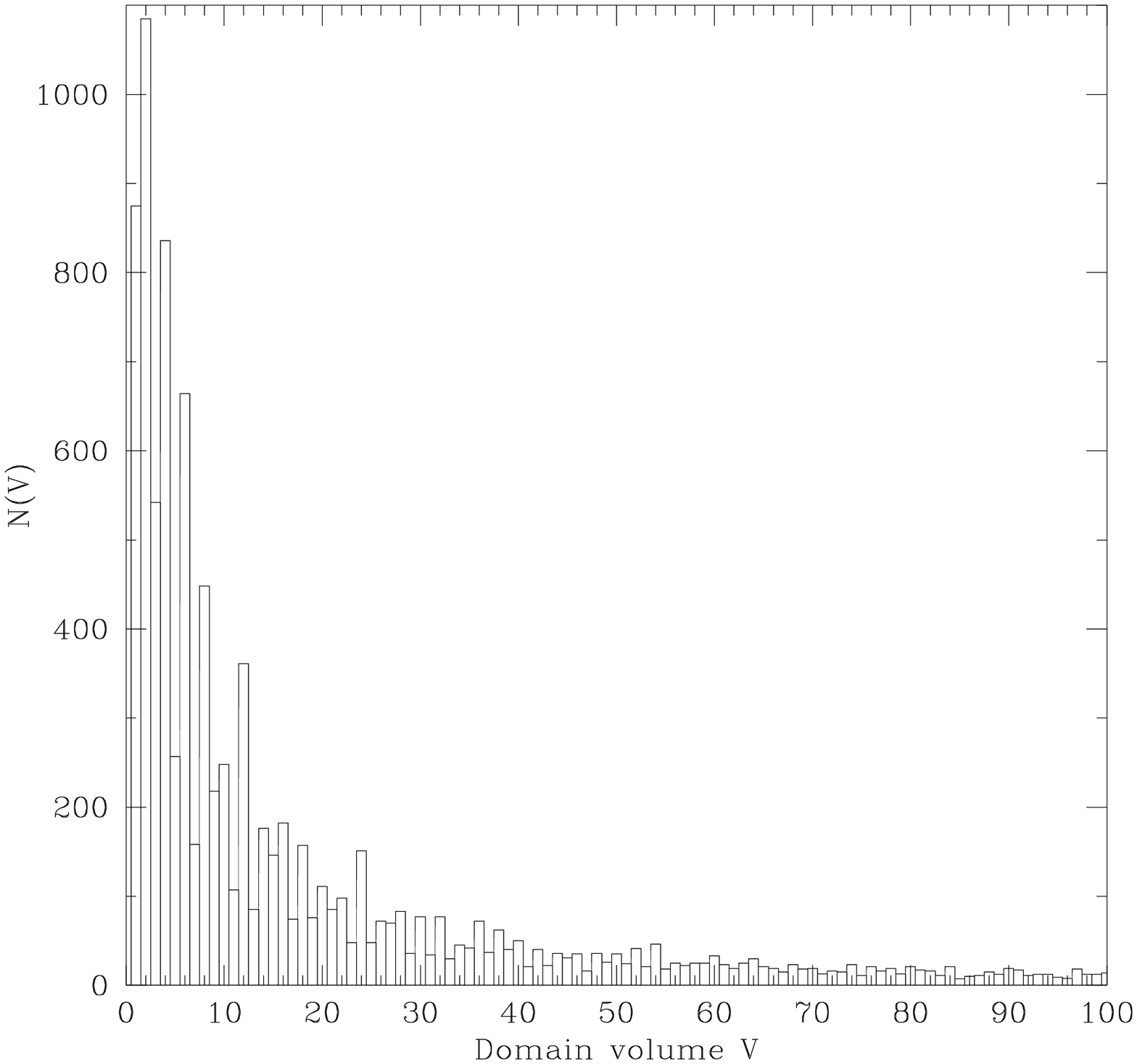,width=2in}}
\put(1.5,0.20){(b)}
\put(0.5,2.6){\psfig{file=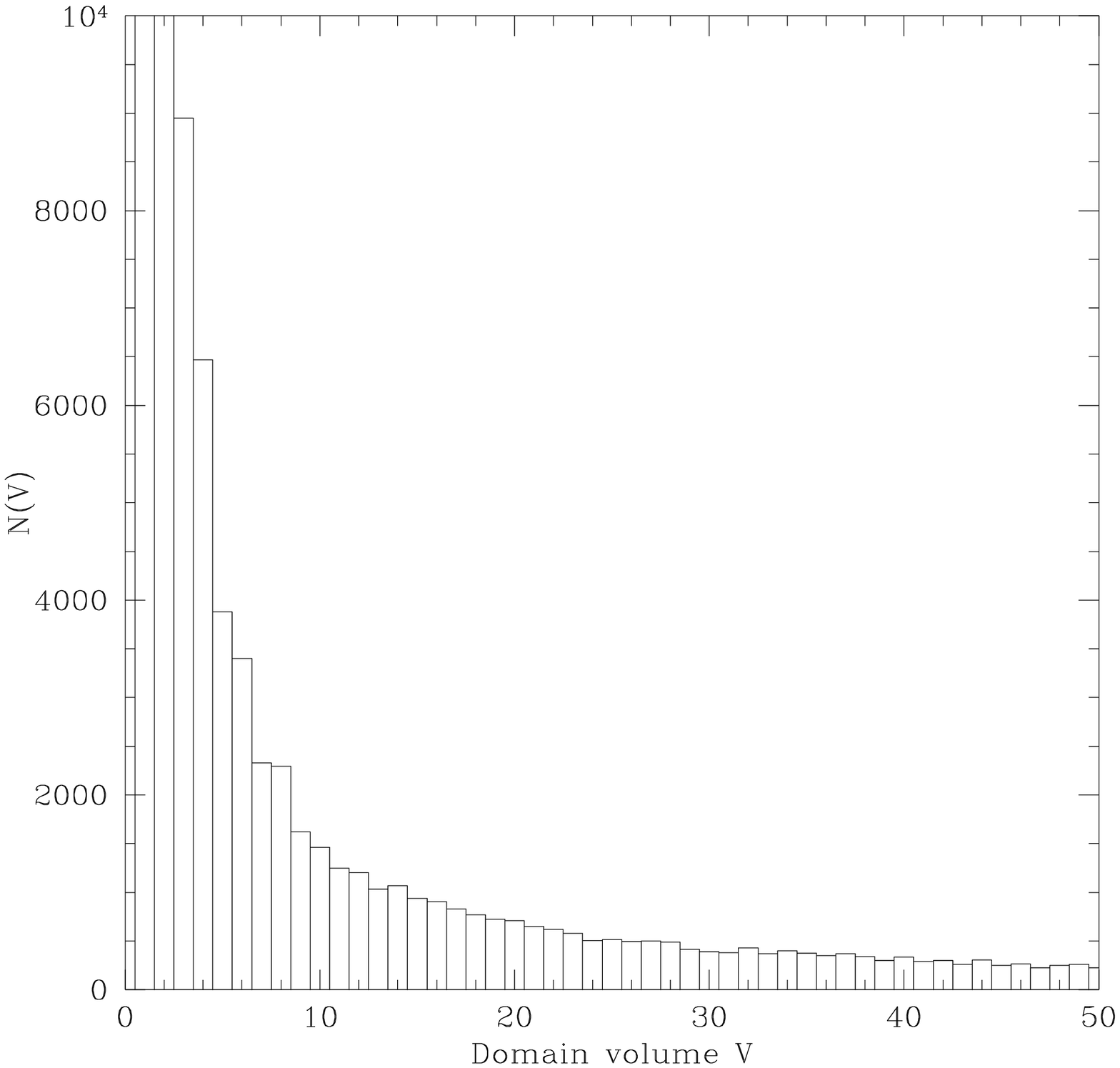,width=2in}}
\put(1.5,2.4){(a)}
\end{picture}
\end{minipage}
\caption{Histograms of domain volumes.
Each illustrates range of domain sizes present for a particular
realisation of the laying algorithm, which fills the box with domains
of diameter randomly chosen in the range 1 to $D$. Figure (a) shows the
results for a $100^3$ box with $D$=5; (b) $D$=15.}
\end{figure*}

\section{Results}
\begin{figure}
\centerline{\psfig{file=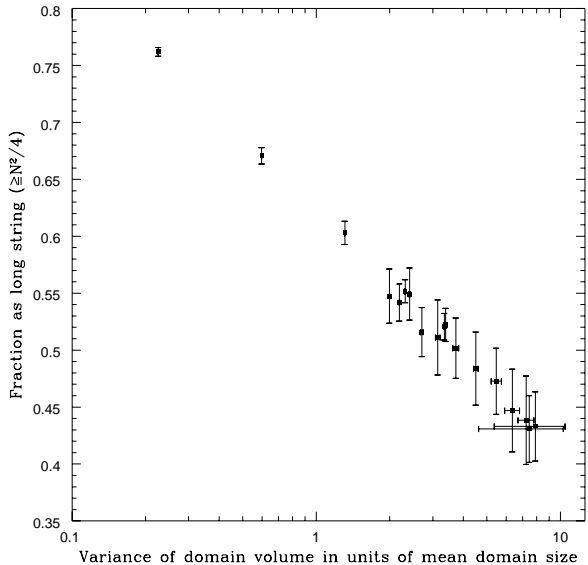,width=3.2in}}
\caption{Results for $100^3$ lattice.}
\label{fig:main}
\end{figure}

In figure \ref{fig:main}, each point is the average of 20
runs, with fixed limits on the range of sizes of domains laid down.
The first point is the result of filling half the box with domains of
side 1 or 2, randomly chosen. The remaining space is filled with unit
domains, in order to achieve a low volume variance. Subsequent points
correspond to the box being filled entirely with domains of sides
between 1 and $D$ ($D$=2,3,...,18).

%%%%%%%%%%%%%%%%%%%%%%%%%%%%%%%%%%%%%%%%%%%%%%%%%%
% Main plot here
%%%%%%%%%%%%%%%%%%%%%%%%%%%%%%%%%%%%%%%%%%%%%%%%%%

The initial decrease in $f_{\infty}$ with increasing variance is
perhaps intuitively understood. Typically, on a regular lattice, the
string performs a self-avoiding random walk - one in which the string
is not permitted to intersect itself or any other strings except at
the `origin'. This property is imposed on the strings by the
restrictions of the lattice method. Such a walk has the property that
the end-to-end distance $l$, in units of the step length,
is related to the mean displacement $R$ by
\[
R \sim l^{3/5}.
\]
In fact, the presence of other
strings provides an extra repulsive effect and so gives the string
near-Brownian characteristics $(R \sim l^{1/2})$. This is confirmed
in the simulations --- typical figures for the $l$ exponent were
$\simeq 0.47 \pm 0.04$ at zero variance.
However, the presence of extended
regions of space from which the string is excluded, i.e. larger
domains, provides restrictions on the ability of the string to `fold
in' on itself. Effectively, in the region of these larger domains, we
expect that a loop of a given radius will have a smaller perimeter
than a similar loop in a region of unit domains. This will increase
the density of loops below the cutoff size.

It is worth noting that as the variance increases,
there exist more ways to fill the box and so
a wider range of possible domain configurations. This also emphasises
the point that volume variance is almost certainly
not the only parameter describing
the spatial distribution of phases that determines $f_{\infty}$.
Statistics at high values of $N$ became unreliable, but it is intriguing
to speculate whether further increases in the variance could well
force all string to be in the form of small loops. The finite size of
the simulation limits the maximum value of N we can reasonably investigate.

\section{Percolation effects}
Another way to observe a reduction in the density of infinite string
is to impose a `tilt' on the vacuum manifold, statistically favouring
the occurrence of one phase \cite{Tan}.

%%%%%%%%%%%%%%%%%%%%%%%%%%%%%%%%%%%%%%%%%%%%%%%%%%
% Tilted vacuum plot here (fig. 3)
%%%%%%%%%%%%%%%%%%%%%%%%%%%%%%%%%%%%%%%%%%%%%%%%%%
\begin{figure}
\centerline{\psfig{file=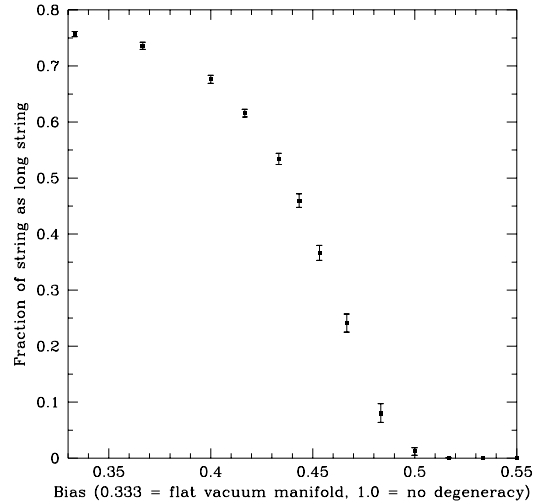,width=2.8in}}
\caption{Plot of $f_{\infty}$ against the bias parameter $\gamma$ for
a zero-variance, $100^3$ box. Each point is averaged over 20 realisations.}
\label{fig:perc}
\end{figure}

If we employ a three-point discretisation
and gradually increase a bias parameter $\gamma$, such that $ P({\mbox{phase
1}}) = \gamma, P(2)=P(3)= \frac{1}{2}(1-\gamma) $, we find that
$f_{\infty}$ drops smoothly, reaching zero at $\gamma \simeq 0.5$
(Fig.\ \ref{fig:perc}).
A phase is said to percolate when it is possible to trace continuous
`infinite' paths through that phase in the box.

Clearly there {\it is} a relation between the strings percolating
(passing through the Hagedorn transition)
and the percolation of phases in the box in this minimally discretised
case. For a string to exist it must have all three phases around it.
An infinite string will therefore ensure that all phases percolate
(including diagonally adjacent regions).
At least one phase percolates for all values of $\gamma$ ---
the critical probability for the occurrence of one phase $p_c$,
above which it percolates, is 0.31 \cite{VV}.
In fact, we note that when $\gamma \simeq
0.5$, $P(2) = P(3) \simeq 0.25$, which is very near the percolation
threshold.

It is interesting to ask whether there is a connection between
increasing the variance of the domain volume and moving away from
string percolation.
Statistical fluctuations in the volume of the
domains will result in the fractions of box volume occupied by each
phase departing from $1/3$, becoming more divergent as the range of
sizes of domains increases. We suggest that this can be interpreted as
an effective tilt of the vacuum manifold.

The fact that a bias will reduce the amount of long string
is easily understood. We consider the probability $p$ that a given
plaquette is pierced by a string ($p=8/27$ ($\approx 0.30$)
in the case of three-point
discretisation). Given that we have an ingoing string through one
face, what is the probability that this string will turn through $\pi
/2$ in the cell under consideration? Obviously the opposite face has
four independent phases (1,2 or 3) attached to it, so the probability
of it containing an {\it outgoing} string is $p/2$. Given this
configuration, the probability of the cell containing a further
ingoing/outgoing string is 1/4. Thus the probability that our string
continues through the cell undeviated, assuming we pair strings
within the box randomly, is
\[
\left( \frac{p}{2} \times \frac{3}{4} \right) +
\left( \frac{p}{2} \times \frac{1}{4} \right)
\times \frac{1}{2} = \frac{7}{16}p.
\]
As we increase the bias parameter $\gamma$ we obviously decrease the
probability $p$. In fact,
\[ p = 2 \gamma (1- \gamma)^2,\]
giving values for $p$ of 0.30, 0.15 and 0.02 for $\gamma = 1/3$,
2/3 and 8/9 respectively.
Thus, strings will be more likely to fold up as $\gamma$
grows, and the population of small loops will increase.
This agrees with ref. \cite{HS}, who point out that
as the bias increases and the strings stop percolating, the fractal
dimension of the strings becomes higher than two and they tend to
`crumple up' more --- they become self-seeking random walks.
Unfortunately,
statistics were too poor to investigate any change in the fractal
dimension of the strings as the bias or the variance was increased.
\begin{figure}
\centerline{\psfig{file=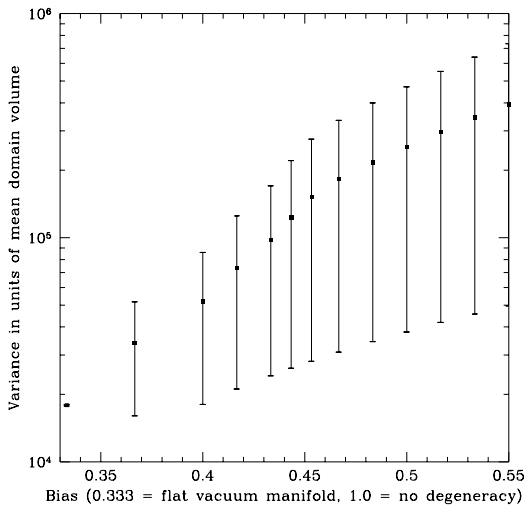,width=2.8in}}
\caption{Correlating the volume variance with the bias parameter $\gamma$}
\label{fig:correlate}
\end{figure}
However, it is possible to investigate qualitatively the connection
between domain volume variance and a vacuum tilt by calculating values for
the variance for each value of the bias
$\gamma$ in figure \ref{fig:perc}.
We set up the box by throwing down phases in single
cells according to the biased probability distribution. We then group
adjoining cells containing the same phase to form larger domains,
whose volumes we calculate.
The results are plotted in figure \ref{fig:correlate}.
The errorbars are misleading
since there is clearly a correlation, and this is to be expected
intuitively --- the more one phase appears at the expense of others, the
bigger the range of sizes of domains present.

\begin{figure}
\centerline{\psfig{file=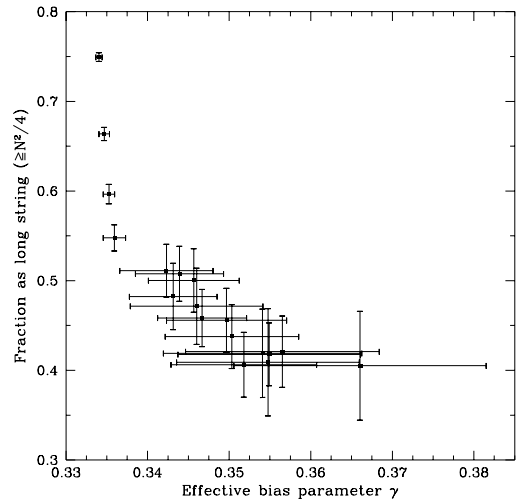,width=2.8in}}
\caption{Calculation of the effective bias parameter in the
minimally-discretised case. For increasing values of N, a value for
$\gamma$ was calculated from the fractions of the box occupied by each
of the three phases.}
\label{fig:effective_tilt}
\end{figure}

Exploring the idea further, we calculate the bias parameter `geometrically',
given the volume occupied by each of the three phases in the box. The
results of this procedure are shown in figure
\ref{fig:effective_tilt}. The values of $\gamma_{\mbox{eff}}$ are too
low to correspond to those in the original figure (\ref{fig:perc}), and
the plots are not similar in form. However, the increase of
$\gamma_{\mbox{eff}}$ as $f_\infty$ decreases is in agreement.

%%%%%%%%%%%%%%%%%%%%%%%%%%%%%%%%%%%%%%%%%%%
% Fig. 5 here
%%%%%%%%%%%%%%%%%%%%%%%%%%%%%%%%%%%%%%%%%%%

\section{Conclusions}
We have seen that a simple extension to the accepted numerical model
for string formation can yield a significantly different estimate of
the amount of infinite string present. It seems feasible that
increasing the variance of the volumes of regions with different VEVs
is equivalent to an effective tilt of the vacuum manifold, that leads
to a reduction in the density of infinite string when considering a
finite volume with periodic boundary conditions. Whether this is the
case in the infinite-volume limit is more debatable. With three-point
discretisation, the variance
becomes ill-defined in this regime, as all three phases percolate.
However, in this
limit it is also unclear whether there is truly a population of
infinite string, distinct from the $l^{-5/2}$ loop distribution, or if
it is purely an artefact of the boundary conditions.

As yet there is no physical argument to suggest what the volume
variance in a given phase transition will be --- and even, considering
the effects of phase equilibration at domain boundaries, how
well-defined this quantity is. Models of dynamic defect formation, even
with simplified treatments of the physics involved in a real
phase transition, may give improved predictions of the defect
configurations \cite{Julian}.
%Preliminary results suggest that the amount of long string is strongly
%suppressed and may even be zero.

One of the consequences of the existence of GUT-scale strings is the
possibility of their being responsible for structure formation.
It is only the infinite string and large loops that will survive long
enough to be useful in this scenario, as
a huge number of Hubble times elapse between string formation and
when perturbations on interesting
(galactic) scales will begin to grow. It may be that if the amount of
long string present is very low, then their structure-seeding
properties will be less significant than previously thought.
Certainly, the proposed existence of a unique scaling solution for the
string network, independent of initial conditions, would be put into
doubt.

\section*{Acknowledgements}
The authors would like to thank Pedro Ferreira and Julian Borrill for
helpful discussions, and James Robinson for contributing part of the
code. A.Y. was funded by PPARC.

\end{document}